\documentclass[sigconf]{acmart}
\AtBeginDocument{%
  \providecommand\BibTeX{{%
    \normalfont B\kern-0.5em{\scshape i\kern-0.25em b}\kern-0.8em\TeX}}}

\copyrightyear{2024}  
\acmYear{2024}  
\setcopyright{rightsretained}  
\acmConference[KDD '24]{Proceedings of the 30th ACM SIGKDD Conference on Knowledge Discovery and Data Mining}{August 25--29, 2024}{Barcelona, Spain} \acmBooktitle{Proceedings of the 30th ACM SIGKDD Conference on Knowledge 
Discovery and Data Mining (KDD '24), August 25--29, 2024, Barcelona, Spain}\acmDOI{10.1145/3637528.3671600}
\acmISBN{979-8-4007-0490-1/24/08}

\usepackage{comment}
\usepackage[caption=false]{subfig}
\usepackage{graphicx}
\usepackage{textcomp}
\usepackage{xcolor}
\usepackage{url}
\usepackage{tabularx}
\usepackage{colortbl}
\usepackage{geometry}
\usepackage{multirow}
\usepackage{caption}
\usepackage{booktabs}
\usepackage{lipsum}
\usepackage{colortbl} 

\acmSubmissionID{ads0493}
\begin{document}
    
\title{EEG2Rep: Enhancing Self-supervised EEG Representation Through Informative Masked Inputs}

\author[1]{Navid Mohammadi Foumani}
\authornotemark[1]
\orcid{0000-0003-2475-6040}
\affiliation{%
  \institution{Monash~University}
  \city{Melbourne}
  \country{Australia}
}
\email{navid.foumani@monash.edu}
\author{Geoffrey Mackellar}
\affiliation{%
  \institution{Emotiv Research}
  \city{Sydney}
  \country{Australia}
}
\email{geoff@emotiv.com}
\author{Soheila Ghane}
\affiliation{%
  \institution{Emotiv Research}
  \city{Melbourne}
  \country{Australia}
}
\email{soheila@emotiv.com}
\author{Saad Irtza}
\affiliation{%
  \institution{Emotiv Research}
  \city{Sydney}
  \country{Australia}
}
\email{saad@emotiv.com}
\author{Nam Nguyen}
\affiliation{%
  \institution{Emotiv Research}
  \city{Sydney}
  \country{Australia}
}
\email{namnguyen@emotiv.com}
\author{Mahsa Salehi}
\affiliation{%
  \institution{Monash~University}
  \city{Melbourne}
  \country{Australia}
}
\email{mahsa.salehi@monash.edu}
\renewcommand{\shortauthors}{N. M. Foumani and et al.}

\begin{abstract}
Self-supervised approaches for electroencephalography (EEG) representation learning face three specific challenges inherent to EEG data: (1) The low signal-to-noise ratio which challenges the quality of the representation learned, (2) The wide range of amplitudes from very small to relatively large due to factors such as the inter-subject variability, risks the models to be dominated by higher amplitude ranges, and (3) The absence of explicit segmentation in the continuous-valued sequences which can result in less informative representations. To address these challenges, we introduce \textit{EEG2Rep}, a self-prediction approach for self-supervised representation learning from EEG. Two core novel components of EEG2Rep are as follows: 1) Instead of learning to predict the masked input from raw EEG, EEG2Rep learns to predict masked input in latent representation space, and 2) Instead of conventional masking methods, EEG2Rep uses a new semantic subsequence preserving (SSP) method which provides informative masked inputs to guide EEG2Rep to generate rich semantic representations. In experiments on 6 diverse EEG tasks with subject variability, EEG2Rep significantly outperforms state-of-the-art methods. We show that our semantic subsequence preserving improves the existing masking methods in self-prediction literature and find that preserving 50\% of EEG recordings will result in the most accurate results on all 6 tasks on average. Finally, we show that EEG2Rep is robust to noise addressing a significant challenge that exists in EEG data. Models and code are available at:\url{https://github.com/Navidfoumani/EEG2Rep}


\end{abstract}


\received{8 February 2024}

\maketitle
\section{Introduction}

\begin{figure}
\vspace{-0.3cm}
  \centering
  \subfloat[MAEEG \textnormal{(Acc:75.21\%)}]{
    \includegraphics[trim=1cm 0.5cm 3.5cm 0cm, width=0.35\linewidth]{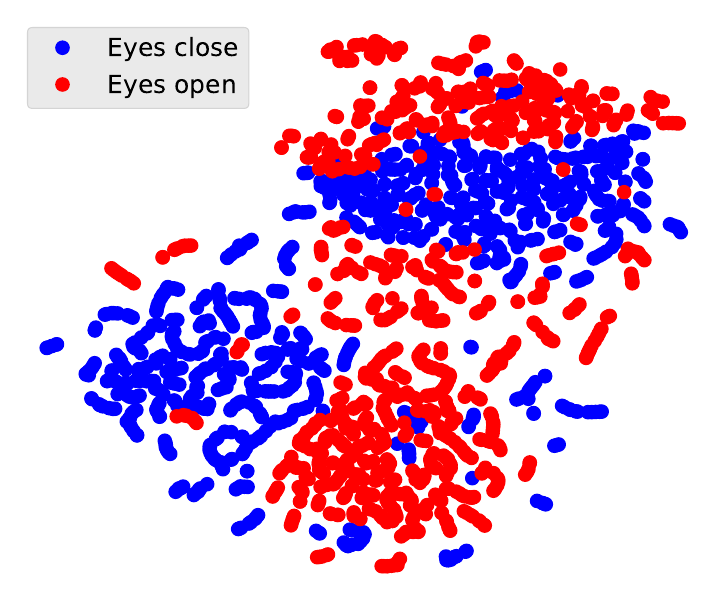}
    \label{fig:SMAEEG-tSNE}%
  }\hfill%
  \subfloat[EEG2Rep \textnormal{(Acc: 81.66\%)}]{%
  \centering
    \includegraphics[trim=3.5cm 0.5cm 1cm 0cm, width=0.35\linewidth]{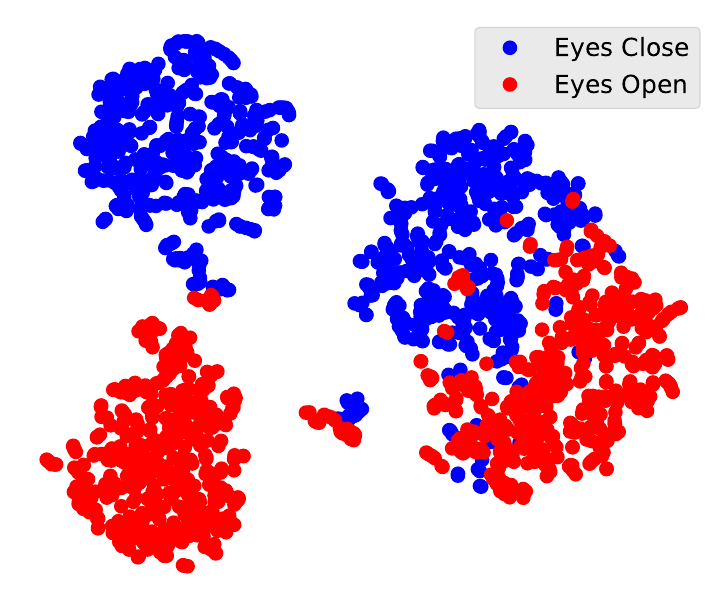}
    \label{fig:our-tSNE}%
  }
  \caption{Comparison of 2D t-SNE plots for representation learned by (a) MAEEG and (b) EEG2Rep on the Crowdsourced EEG dataset.}
  \label{fig:tSNE}
  \vspace{-0.6cm}
\end{figure}

An electroencephalogram (EEG) is a noninvasive method that captures brain data by placing electrodes on the patient's scalp surface, enabling the recording of electrical activity within the brain~\cite{EEG2002}. This specialized and complex biological electrical signal serves as a reflection of the brain's functional state, providing insights into the individual's mental condition~\cite{EEG2005}. From this data, valuable information can be extracted, including vital signs that facilitate continuous monitoring of the patient's health~\cite{lotte2018review}. Additionally, EEG plays a crucial role in diagnosing and identifying various brain conditions, finding applications in diverse healthcare domains such as sleep medicine, neurological disorders, cardiovascular disease detection, and activity monitoring~\cite{roy2019deep,chen2022brainnet,rakovic2024measuring}.

In the past decade, the integration of deep learning into biomedical research has experienced significant growth, demonstrating its ability to frequently outperform conventional machine learning methods across various tasks~\cite{roy2019deep,craik2019deep,hosseini2020review,rakovic2024measuring,weng2024self}. 
However, the training of deep learning models for biomedical applications requires substantial amounts of data, annotated by experts, whose collection is often time and cost-prohibitive. Furthermore, deeper neural networks are susceptible to overfitting the EEG data, particularly in the presence of inter-subject variability~\cite{Bendr,schirrmeister2017deep}. Self-supervised learning (SSL) has emerged as a prominent solution for such problems, as it allows learning powerful representations from vast unlabeled data by producing supervisory signals directly from the data~\cite{Bendr,maeeg,yang2023biot,weng2024self}.


In EEG analysis, two commonly used self-supervised learning approaches are invariance-based methods and self-prediction methods~\cite{foumani2023deep,weng2024self}. In invariance-based methods, the objective is to optimize an encoder to generate similar embeddings for two or more views of the same EEG time series~\cite{foumani2023deep}. These different views of the time series are usually crafted using a set of manual data augmentations, including techniques like jittering, permutation, and scaling~\cite{TS-TCC2021,MohsenvandIM20}. The main idea behind self-prediction methods is to remove or corrupt parts of the input and train the model to predict or reconstruct the altered content~\cite{foumani2023deep}. For example, approaches like Masked AutoEncoders (MAE)~\cite{mae2022} learn representations by reconstructing randomly masked patches from an input~\cite{Bendr,mae2022,TST2021}.

Invariance-based pretraining methods can construct representations of high-level semantics by capturing essential features consistent across various data views. This strategy enables the model to identify and prioritize features important for understanding the underlying semantics of the data, as these features remain unchanged despite variations in the input.
However, we believe that invariance-based methods perform well in computer vision and natural language processing due to the strong constraints present in image and text data. For example, the success with images arises from tasks related to object interpretation, where transformations such as scaling, blurring, and rotation assume that the resulting images will be similar to those generated in the original scenario with changes in camera zoom, stability, focus, or angle. However, there do not appear to be equivalent transformations that can be applied to EEG data. Augmentation methods applied to EEG data can inadvertently modify the semantic meaning, underlying distribution, and class representation of the signals~\cite{ling2022staging,series2vec23}. For example, augmentation may introduce synthetic patterns or artifacts that do not align with genuine brain activity.

Moreover, invariance-based methods may introduce significant biases, potentially hindering specific downstream tasks or even pretraining tasks with diverse data distributions. The generalization of these biases for tasks requiring varying levels of abstraction often remains unclear. Distinct data augmentation strategies may result in misinterpretation or misclassification by the model. For example, in scenarios such as sleep stage classification (involving low-frequency bands) and emotion recognition (involving high-frequency bands), identical augmentations may not be suitable for both cases.

In contrast to invariance-based methods, self-prediction pretraining tasks demand less prior knowledge and demonstrate ease of generalization across diverse downstream tasks~\cite{data2vec2022,data2vec2-2023}. However, self-prediction pretraining faces unique challenges when applied to EEG data \cite{maeeg, Bendr} which makes it ineffective. 
Fig.~\ref{fig:SMAEEG-tSNE} depicts the visualization of EEG representations by a state-of-the-art self-prediction EEG pretraining model, namely MAEEG \cite{maeeg}, on Crowdsourced dataset~\cite{crowdsourced} with two classes. The two classes are not easily separable in the learned representation space, resulting in low classification accuracy. Here, we outline the three main challenges that exist in EEG data:

\begin{itemize}
    \item \textbf{Challenge 1:} The recorded EEG is invariably contaminated with noise, impacting the reconstruction loss function and potentially introducing significant errors. Even accurate predictions may yield high errors due to the pronounced impact of noise on the loss function. 
    
    \item \textbf{Challenge 2:} EEG data has a wide range of amplitude values, which can be due to the variability between different subjects or the variability of electrode placement. The substantial ranges make it particularly challenging to reconstruct accurate values during the reconstruction process. 
    
    \item \textbf{Challenge 3:} EEG signals differ from text and images in that there is no explicit segmentation for EEG data, as they are continuous-valued sequences.
    Hence, the implementation of a masking strategy becomes essential for a model to have sufficient enough information for the reconstruction of the masked EEG.
\end{itemize} 

To address the challenges mentioned above, we introduce EEG2Rep to enhance the semantic quality of EEG representations without relying on prior knowledge about downstream tasks.
In contrast to the existing self-prediction methods that learn EEG representations by reconstructing the raw EEG data space \cite{maeeg,Bendr}, EEG2Rep is trained to reconstruct more abstract features of EEG data in the latent space. Such approaches have shown to be effective in image and text representation learning \cite{data2vec2-2023,i-jepa23,zhou2022image}. Our motivation is that the existing noise in EEG is less likely to remain in the abstract features of EEG, and by learning to reconstruct the abstract features we potentially eliminate the unnecessary noise that exists in raw EEG data (addressing challenge 1). 
Additionally, as the abstract features are normalized within the representation space, the reconstruction of these features becomes more straightforward compared to reconstructing the potentially high amplitude value range of raw EEG data (addressing challenge 2).
Finally, EEG2Rep leverages our novel \textit{Semantic Subsequence Preserving} method to ensure that the context has sufficiently meaningful and rich semantic information (addressing challenge 3). 

Fig.~\ref{fig:our-tSNE} displays the visualization of EEG representations learned by EEG2Rep. The two classes in this figure are easily separable which highlights the effectiveness of EEG2Rep in enhancing EEG representations and, consequently, improving classification accuracy.
Another core component of EEG2Rep is its efficient multi-masking design. Specifically, we reuse the same target representation for various masked versions of each sample. Additionally, we predict the representation of various target blocks for a single masked input to improve efficiency further. 

This work follows from the project with Emotiv Research~\footnote{\href{https://www.emotiv.com/neuroscience-research-education-solutions/}{www.emotiv.com/research}}, a bioinformatics research company based in Australia, and Emotiv, a global technology company specializing in the development and manufacturing of wearable EEG products. In our prior work, we looked at detecting distraction episodes in drivers by analyzing their brain EEG as a case study. One significant challenge we encountered was the presence of noise in the recorded EEG datasets and the inability of current supervised detection models to learn patterns of distraction within the presence of noise. This paper addresses this issue, along with the two additional challenges mentioned above.


\section{Methodology of EEG2Rep}

\begin{figure*}
    \centering
    \includegraphics[trim=1cm 7cm 6cm 0.5cm, width=0.85\textwidth]{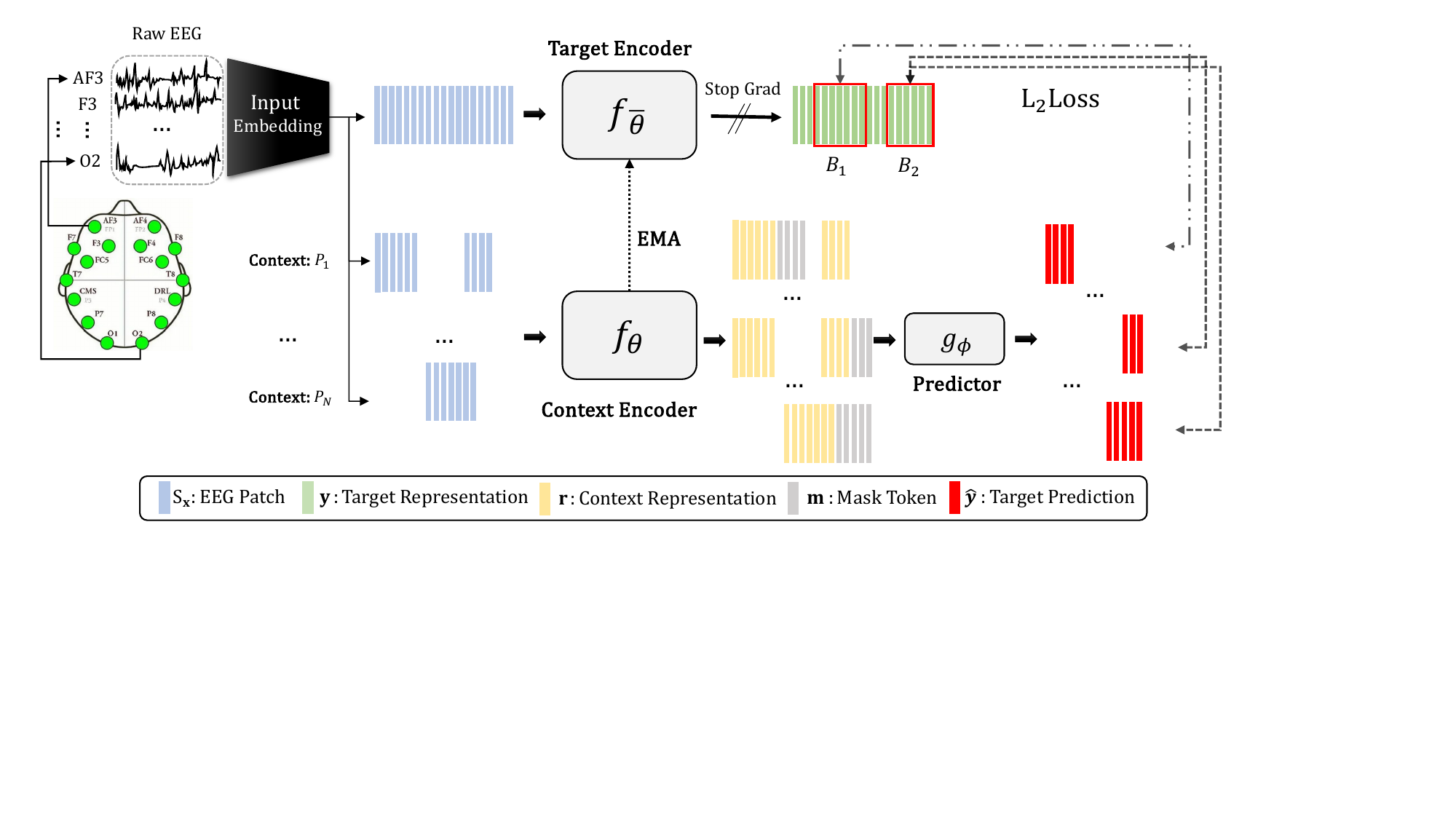}
    \caption{Architecture of EEG2Rep }
    \label{fig:EEG2Rep}
\end{figure*} 
\subsection{Problem Definition}   
Our goal is to address the problem of learning a nonlinear embedding function that can effectively map each EEG sample $X_i=\{x_1,x_2,\ldots,x_L\}$ from a given dataset $D$ into a concise and meaningful representation $R_i \in \mathbb{R}^{d_e}$, where $d_e$ indicates the desired representation dimension. The EEG dataset $D$ consists of $n$ samples, denoted as $D=\{{X_1},{X_2},...,{X_n}\}$, with each ${X_i}$ representing a $C$-channel EEG sequence of length $L$. To evaluate the quality of our learned representation $\mathbf{R}=\{R_1, R_2,\ldots, R_n\}$, we examine two scenarios based on the availability of labeled data: i) \textit{Linear Probing}: We first pre-train a model without labels through a self-supervised pretext task. Upon completing the pre-training phase, we freeze the encoder and add a linear classifier on top of the pre-trained model's output or intermediate representations. The linear classifier is then trained on a downstream task, typically a classification task, utilizing the pre-trained representations as inputs, and ii) \textit{Fine-Tuning}: Initially, we pre-train a model without labels through a self-supervised pretext task. Next, we perform fine-tuning by training the entire model for a few epochs using a labeled dataset in a fully supervised manner.
\vspace{-0.2cm}
\subsection{Model architecture}
Illustrated in Fig.~\ref{fig:EEG2Rep}, the EEG2Rep model is introduced with the main goal of predicting the representation of a given EEG sample based on a masked view of the same EEG input. We now explain each component of this architecture separately in the following subsections.
\subsubsection*{\textbf{Input Embedding}}\label{Input-Embedding}
Building upon established works~\cite{schirrmeister2017deep,chambon2018deep,foumani2021disjoint,TS-TCC2021,ConvTran2023,series2vec23}, we adopt a 3-layer convolutional neural network as input embedding to convert the raw EEG data into patches. Specifically, we feed EEG sample $X_i$ to the first layer that incorporates a depthwise convolutional layer, specifically designed to capture the spatial correlations between channels~\cite{schirrmeister2017deep,foumani2021disjoint}. This is succeeded by a linear spatial filter to amplify the signal-to-noise ratio. The spatial filters leverage the fact that neural signals exhibit specific spatial patterns across the scalp, while noise sources may manifest more random spatial patterns~\cite{chambon2018deep}. 

Following the spatial filter, we integrate max pooling and spatial padding to ensure translation equivalence and effectively address edge effects. The output of this network is a set of EEG patches $\hat{S}_\textbf{x} = \{\hat{S}_{x_1}, ..., \hat{S}_{x_l}\}$ where $\hat{S}_{x_i} \in \mathbb{R}^{d_x}$, and $d_x$ is the embedding dimension, and $l$ is the number of patches. Lastly, for every patch $\hat{S}_{x_i}$, the positional embedding feature of the $i^{th}$ position is added to it, resulting in the EEG patches $S_\textbf{x} = \{S_{x_1}, ..., S_{x_l}\}$ (shown in blue in Fig.~\ref{fig:EEG2Rep}). Please note for each EEG sequence $X_i$, we will have a set of EEG patches $S_\textbf{x}^i$ as the output of the input embedding network. However, for simplicity, we will drop the superscript $i$ from $S_\textbf{x}^i$ and use $S_\textbf{x}$ as an input to the following components in the EEG2Rep architecture. 

\subsubsection*{\textbf{Context-driven target prediction}}
Previous self-prediction methods for EEG data have primarily followed the approach of Masked Autoencoders (MAE) \cite{mae2022}, reconstructing local windows of the raw input EEG~\cite{maeeg}, or have adopted a BERT-like method \cite{bert2018}, predicting discrete representations~\cite{Bendr}. However, the resulting representations often demonstrate lower semantic quality than invariance-based methods during off-the-shelf evaluations, such as linear probing, due to the intrinsic characteristics of EEG data mentioned earlier. Low signal-to-noise ratio in EEG challenges the reconstruction task and a wide range of amplitudes in EEG further complicates the optimization process.

To enhance the semantic depth of self-supervised representation learning, we introduce the concept of \textit{context-driven target prediction} for EEG data. In this approach, the model is trained to predict the representations of the original unmasked training data based on an encoding derived from the masked sample in abstract representation space. Compared to self-prediction methods that predict in raw space, EEG2Rep uses abstract prediction targets for which unnecessary raw-level details or noise are potentially eliminated, thereby leading the model to learn richer semantic features.

EEG2Rep comprises three main components: \textit{Target network} uses the complete (unmasked) input embedding $S_\textbf{x}$ to generate semantically rich representations, allowing each patch to encode knowledge of all others through its self-attention architecture. The resulting output serves as the target for our learning task. \textit{Context network} shares the architecture with the target network, differing only in its use of the masked version of the input embedding to generate representations for visible patches. Our innovative semantic subsequence preserving method ensures that the context network's output contains sufficient semantic information for reconstruction purposes.  We use a standard transformer~\cite{attention2017,bert2018} for both target and context networks. \textit{Predictor network} leverages the output of the context network and randomly chosen masked tokens to regress the targets, i.e., the output of the target network. We enhance contextualized information integration for the predictor network by incorporating cross-attention-based transformers~\cite{bert2018}. Below, we elaborate on the process of creating the target, context, and predictor networks for the training task, providing clarification on how these networks are parameterized.

\vspace{0.15cm}
\noindent\textbf{Target Network}\\
Given an EEG embedding $S_\textbf{x} = \{S_{x_1}, \ldots, S_{x_l}\}$, we feed it through the target-encoder $f_{\bar{\theta}}$ to obtain a corresponding patch-level representation $\textbf{y} = \{y_1, \ldots, y_l\}$:
\vspace{-0.3cm}
\begin{equation}
    \textbf{y} = f_{\bar{\theta}}(S_\textnormal{x})
\end{equation}
where $y_i\in \mathbf{R}^{d_{e}}$ is the representation associated with the $S_{x_i}$ patch and $d_e$ is the transformer's embedding dimension (shown in green in Fig.~\ref{fig:EEG2Rep}). We apply normalization to these representations, preventing the model from collapsing into a constant representation for all time steps and ensuring that values with high amplitude do not dominate the target features.

To obtain the targets for our reconstruction loss, we randomly sample $M$ blocks from the target representation $\textbf{y}$. $B_i$ is the $i^{th}$ masked block in $\textbf{y}$ where $i\in\{1,2,\ldots,M\}$ and $\textbf{y}(i) = \{y_j\}_{j\in B_i}$  (e.g., $B_1$ and $B_2$ is annotated in red in Fig.~\ref{fig:EEG2Rep} and $\textbf{y(1)}$ contains all $\{y_j\}_{j\in B_1}$). 
Note that the target blocks are chosen from the output of the target encoder, not the input of the target encoder. This distinction is crucial to ensure each target representation retains a high semantic level as it encodes the knowledge of all input patches through the self-attention mechanism in Transformers architecture~\cite{data2vec2022,i-jepa23,data2vec2-2023}.

\vspace{0.15cm}
\noindent\textbf{Context Network} \\ 
To obtain the context in EEG2Rep, we initially sample $N$ contexts from the EEG patches $S_{\textbf{x}}$ (refer to Sec.\ref{Sec:SSP} for details on our novel context selection). Only visible context patches are processed via context-encoder model $f_{\theta}$ to obtain context representations:
\begin{equation}
    \textbf{r}(q) = f_{\theta} (P_q) = f_{\theta} (\{S_{x_t}\}_{t\in P_q})
\end{equation} 
Where $P_q$ correspond to the $q^{th}$ context which contains a subset of EEG patches from $S_{\textbf{x}}$ and $\textbf{r}(q) = \{r_t\}_{t\in P_q}$ is its patch level representation (shown in yellow in Fig.~\ref{fig:EEG2Rep}). Since the target blocks are sampled independently from the contexts, there may be significant overlap. We exclude any overlapping regions from the target block in the loss calculation to ensure a non-trivial prediction task. Examples of various contexts ($P_1$ and $P_N$) and target blocks ($B_1$ and $B_2$) are illustrated in Fig.~\ref{fig:EEG2Rep}. The remaining patches are called masked tokens (shown in grey in Fig.~\ref{fig:EEG2Rep}).

\vspace{0.15cm}
\noindent\textbf{Predictor Network} \\
For the Predictor network, we use a 4-layer cross-attention transformer to enhance the effective correlation among the context representation and masked tokens. The ``cross-attention" aspect of the transformer is capable of blending two distinct embedding sequences, which may vary in length and originate from different sources~\cite {attention2017}. In this setup, the masked token serves as a query, while the key and values are sourced from the context encoder.

Given the output of the context encoder $\textbf{r}(q)$, our goal is to predict $M$ target block representations $\{\textbf{y}(1),\ldots,\textbf{y}(M)\}$. For each target block $\textbf{y}(i)$, the predictor $g_\phi$ takes as input the output of the context encoder $\textbf{r}(q)$ and a mask token $m(i)$ for each patch we wish to predict and outputs a patch-level prediction:
\begin{equation}
    \hat{\textbf{y}}(i) = g_\phi(\textbf{r}(q), m(i))
\end{equation}

where $m(i) = \{m_j\}_{j\in B_i}$. The mask tokens are parameterized by a shared learnable vector with an added positional embedding. Since we wish to make predictions for $M$ target blocks, we apply our predictor $M$ times, each time conditioning on the mask tokens corresponding to the target block locations we wish to predict. 

\textbf{loss}: Given context-driven training targets $\textbf{y}(i)$ and the predicted patch-level representation $\hat{\textbf{y}}(i)$, we use the L2 loss:
\begin{equation}\label{eq:L2}
    Loss_{rec} = \frac{1}{M}\sum_{i=1}^M\sum_{j\in B_i}||y_j-\hat{y}_j||_2^2
\end{equation}

\vspace{0.15cm}
\noindent\textbf{EEG2Rep Weights}\\
The predictor parameters $\phi$ and the context network parameters $\theta$ are optimized through gradient-based methods. However, the target network's parameters $\bar{\theta}$ undergoes updates using an exponentially moving average (EMA)~\cite{Ibot2021,data2vec2022} of the context network parameters:
\begin{equation}
    \bar{\theta} = \tau\bar{\theta} + (1-\tau)\theta
\end{equation}
We implement a schedule for the hyperparameter $\tau$ that linearly increases from an initial value $\tau_0$ to the target value $\tau_e$ during the first $\tau_n$ updates. After this initial phase, the value remains constant for the rest of the training. This approach ensures that the target network is updated more frequently in the early stages of training when the context network is random, and less frequently in later stages when more robust parameters have been learned.

\subsection{Semantic Subsequence Preserving (SSP)} \label{Sec:SSP}
\begin{figure}
    \centering
    \includegraphics[trim=2cm 5.5cm 18cm 0cm, width=0.35\textwidth]{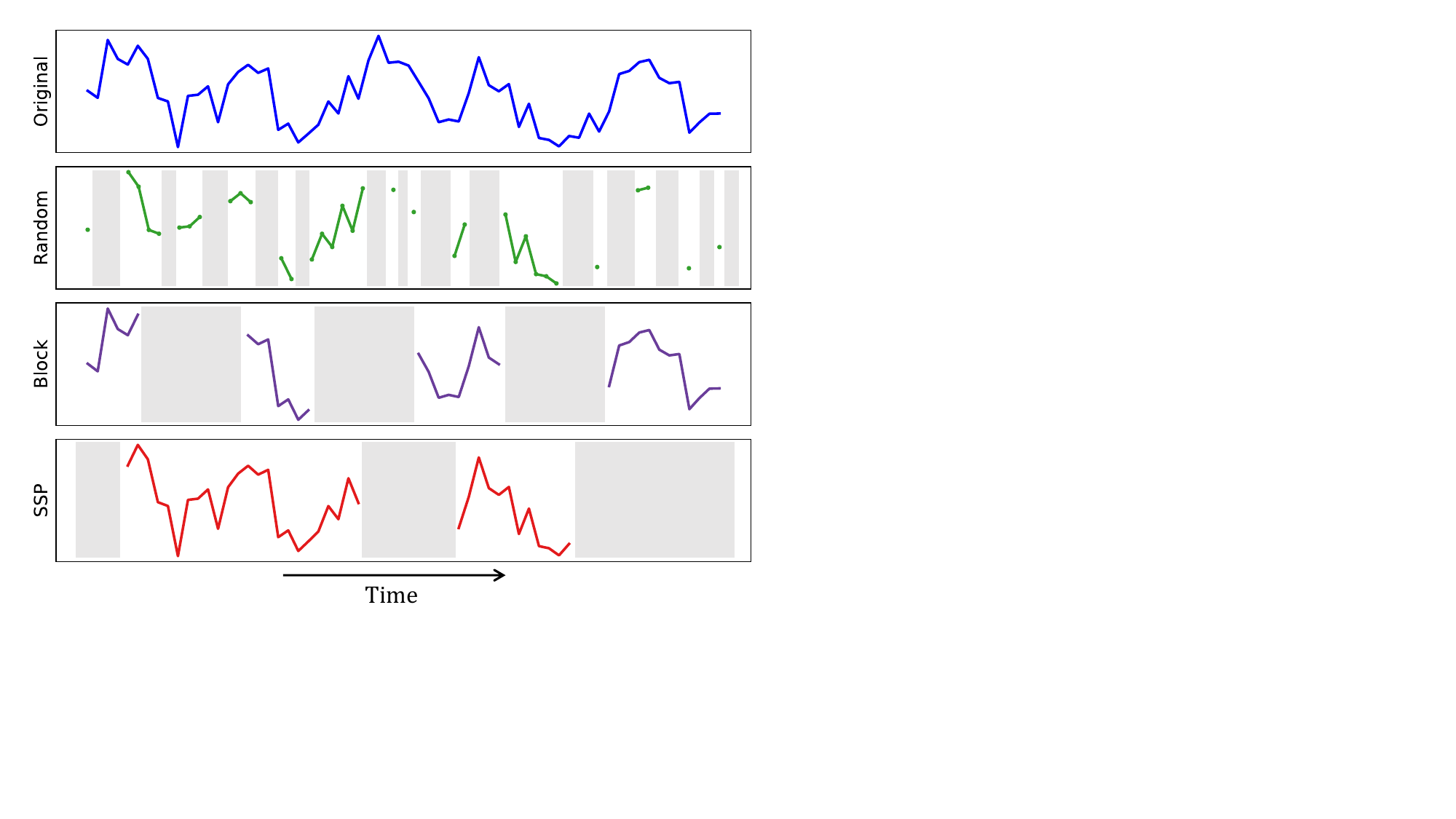}
    \caption{Different masking strategies applied to EEG samples from Crowdsourced datasets~\cite{crowdsourced} with a 50\% masking ratio. From top to bottom: (1) \textit{Original} EEG sample, (2) \textit{Random} masking, (3) \textit{Block} masking, and (4) Semantic Subsequence Preserving (\textit{SSP}).}
    \label{fig:Masking}
    \vspace{-0.5cm}
\end{figure} 

Random masking has proven successful for Masked Autoencoders \cite{mae2022}, where random patches are sampled without replacement, following a uniform distribution, resulting in significant efficiency improvements. Fig.~\ref{fig:Masking} illustrates various masking strategies applied to EEG samples, with the top subplot showcasing the original EEG sample from Crowdsourced datasets~\cite{crowdsourced} recorded for 0.5 seconds at 128 Hz. The subsequent subplots demonstrate different masking techniques, each with a $50\%$ masking ratio. The second subplot exhibits random masking, following the Masked Autoencoders style. While the MAE approach has demonstrated success in computer vision tasks, it may pose challenges in building semantic representations for EEG data due to the lack of structure in the created masks (e.g., random binary masking). Notably, most of the masked regions can be reconstructed trivially, for example, through interpolation. On the other hand, block masking~\cite{data2vec2022,i-jepa23}, involves masking entire blocks of time-steps or patches. However, this approach does not guarantee that large continuous portions of the training sample remain unmasked. As shown in Fig.~\ref{fig:Masking}, block masking primarily focuses on the structure of the mask rather than context. Our objective is to enable the context encoder to construct semantically rich representations over local regions of the sample.

To address this, we propose \textit{Semantic Subsequence Preserving (SSP)}. Instead of determining which time steps to mask, this approach decides which time steps to preserve in a block-wise manner. The process begins by randomly selecting $\beta$ starting points and symmetrically expanding them until the block reaches a width given by:
\vspace{-0.2cm}
\begin{equation}
    \text{Block Size} = \lceil{\frac{(1-\rho)\times l}{\beta}} \rceil
    \label{eq:masking}
\end{equation}
Here, $l$ represents the total number of time steps (patches) in a training sample, $\beta$ is the desired number of blocks, and $\rho$ is the mask ratio.
We allow visible blocks to overlap, leading to oversized subsequences preserving and some variability in the number of visible time steps for each EEG sample. As shown in Fig.~\ref{fig:Masking}, the first preserved subsequence arises from overlapping two blocks. As we encode only visible time steps, we adopt a simple strategy to maintain uniformity in the number of masked time steps across all samples in a batch. Following the semantic subsequence preserving step, we randomly preserve individual time steps until we reach the targeted number of preserved time steps, calculated as $(1-\rho) \times l$.

\subsection{Efficiency: Multiple masking}
As discussed in earlier sections, our model comprises three main components: the target, context, and predictor networks. The processing of each sample through these networks can be computationally intensive. Additionally, computing activations for the target network is less efficient compared to the others, as it requires processing the entire unmasked EEG input. Hence, to mitigate the computation cost, we implement a two-stage approach with multiple masking and multiple predictions.

\vspace{0.15cm}
\noindent\textbf{Multiple Masking}\
We explore $N$ different masked versions of the training sample and compute the loss with respect to the same target representation. This is feasible because target representations are based on the full unmasked version of the sample. As $N$ increases, the computational overhead of computing target representations becomes negligible.

\noindent\textbf{Multiple Prediction}\
Subsequently, having obtained representations of various masked versions of the same input, we predict the representation of various target blocks for a single masked input to further enhance efficiency. Now, the reconstruction loss in Equation~\ref{eq:L2} is updated to:
\vspace{-0.3cm}
\begin{equation}\label{eq:L2-Updated}
    Loss_{rec^*}= \frac{1}{M\times N}\sum_{q=1}^N \sum_{i=1}^M\sum_{j\in B_i}||y_j-\hat{y}_j||_2^2
\end{equation}
\subsection{Loss Regularisation} \label{sec:LossReg}
A common challenge faced by algorithms generating and predicting their own targets is representation collapse, where the model produces similar representations for all masked segments, making the problem trivial to solve \cite{jing2021understanding,ben2023reverse}. Various strategies have been proposed to address this issue such as contrastive learning\cite{SimCLR2020} for invariance-based methods and stop gradient\cite{grill2020bootstrap,data2vec2022,Ibot2021} and additional clustering \cite{ben2023reverse} for self-prediction methods. As we mentioned earlier, we use the exponentially moving average and stop gradient methods to prevent representation collapse similar to other models in CV and NLP like BYOL~\cite{grill2020bootstrap}, Data2vec~\cite{data2vec2022}, IBOT~\cite{Ibot2021}. However, in our experiment, we found that preventing collapse does not guarantee that the model learns high-quality representations. To further enhance the representation learning process, we add Variance-Invariance-Covariance (VICReg)~\cite{vicreg2022} regularization to our current reconstruction loss. VICReg encourages more variety among the data in the batch by using a hinge loss that limits how much the standard deviation can change. It also involves a covariance loss, which penalizes the off-diagonal elements of the covariance matrix of the representation, encouraging less correlation between features.
The final loss of our model is updated to:
\begin{equation}\label{eq:total_Loss}
    \textit{Total Loss} = \lambda Loss_{rec^*} + \mu v(R) + \gamma c(R)
\end{equation}

Where \textit{Total Loss} is minimized over batches of samples and $v(R)$ and $c(R)$ denotes the variance, and covariance losses on the representations, respectively. $\lambda, \mu, \gamma$ are hyperparameters controlling the balance between these loss components.

\section{Related Work}
Traditional approaches to extract useful features from EEG are surveyed in \cite{subha2010eeg}. In recent years, advanced self-supervised learning methods have been proposed aiming to derive representations from sparsely labeled EEG data. Here, we discuss two primary self-supervised approaches for EEG representation learning: invariance-based and self-prediction, and refer interested readers to \cite{weng2024self,foumani2023deep} for more details.

\vspace{-0.2cm}
\subsection{Invariance-based self-supervised learning}
In the field of EEG, a series of studies have drawn inspiration from the SimCLR~\cite{simclr} framework in computer vision and ALBERT~\cite{albert} in NLP, aiming to generate consistent embeddings for various yet compatible views of the same input~\cite{MohsenvandIM20,TS-TCC2021,TF-C2022,foumani2023deep,weng2024self}. This involves capturing consistent characteristics across different views, where EEG views are typically formed using a set of carefully designed data augmentations. For instance, SeqCLR~\cite{MohsenvandIM20} applies diverse strategies like amplitude scaling, temporal shifting, DC shifting, and band-pass filtering to create different views. Similarly, Time-series Temporal and Contextual Contrasting (TS-TCC)~\cite{TS-TCC2021} use weak and strong augmentations to transform input series into two views, utilizing a temporal contrasting module to learn robust temporal representations. The contrasting contextual module is then built upon the contexts from the temporal contrasting module and aims to maximize similarity among contexts of the same sample while minimizing similarity among contexts of different samples.

Time-Frequency Consistency (TF-C) \cite{TF-C2022} leverages the frequency domain to achieve better representation. It proposes that the time-based and frequency-based representations, learned from the same time series sample, should be more similar to each other in the time-frequency space compared to representations of different time series samples. Kan et al. \cite{meiosis} introduced a new augmentation method called `meiosis' that involves randomly exchanging data segments between samples to create positive correlations, using group-level contrastive learning to distinguish emotional states. Furthermore, Shen et al. \cite{Shen2023} implemented cross-subject contrastive learning. They compared negative pairs from different individuals with positive pairs from the same subjects during similar events. This helps the model generalize EEG representations while reducing variations due to individual differences. While invariance-based methods can generate informative representations, they introduce strong biases that may lead to model misinterpretation or misclassification. Additionally, defining generalizable augmentations is more challenging in EEG data compared to other types of data.

\vspace{-0.2cm}
\subsection{Self-Prediction based self-supervised learning}
The primary goal of self-prediction-based models is to reconstruct corrupted or masked input. Following the success of models like BERT~\cite{bert2018}, BErt-inspired Neural Data Representations (BENDR) \cite{Bendr} utilizes a mask-autoencoder approach to generate representations and minimize reconstruction loss between masked and reconstructed features. BENDR encodes multi-channel EEG signals into temporal embeddings using a 1D convolution block, inspired by the wave2vec approach \cite{wav2vec}. It then employs a transformer encoder to process locally masked EEG signals for feature and representation extraction. Pre-trained on the extensive TUH EEG dataset~\cite{TUAB,TUEV}, the model shows improved performance across tasks like motor imagery, sleep stage classification, and event recognition. Similarly, MAEEG \cite{maeeg} refines this approach by adding two layers to convert the transformer outputs back to the original EEG dimensions, reconstructing the EEG signal from contextual features and minimizing the loss between original and reconstructed signals. Li et al.~\cite{li2022multi} introduce a multi-view mask autoencoder that reconstructs masked EEG content across spectral, spatial, and temporal dimensions to derive emotion-related EEG features. 

In contrast to invariance-based methods, self-prediction pretraining tasks demand less prior knowledge and demonstrate ease of generalization across diverse downstream tasks~\cite{data2vec2022}. However, the resulting representations are typically of a lower semantic level and may underperform invariance-based pretraining in off-the-shelf evaluations like linear probing and pretraining, mainly due to the intrinsic nature of EEG data. In this study, we explore ways to enhance the semantic level of self-supervised representations without relying on additional prior knowledge encoded through data augmentation. To achieve this, we introduce EEG2Rep to improve self-supervised EEG representations through informative masked inputs.
\vspace{-0.4cm}
\begin{table}[ht]
    \centering
    \caption{Overview of EEG Datasets}
    \vspace{-0.4cm}
    \footnotesize
    \setlength{\tabcolsep}{1pt}
    \setlength\extrarowheight{5pt}
    \begin{tabular}{llccccc}
        & \textbf{Datasets} & \textbf{Rate} & \textbf{Dim} & \textbf{Len} & \textbf{\#Samples} & \textbf{Task} \\
        \hline \hline
        \parbox[t]{2mm}{\multirow{4}{*}{\rotatebox[origin=c]{90}{\textbf{Emotiv}}}}
         & DREAMER~\cite{dreamer} & 128Hz & 14 & 2s & 77,910 & Emotion Detection \\
         & STEW~\cite{stew} & 128Hz & 14 & 2s & 26,136 & Mental workload Classification \\
         & Crowdsourced~\cite{crowdsourced} & 128Hz & 14 & 2s & 12,296 & Eyes open/close Detection \\
         & Driver Distraction & 128Hz & 14 & 2s & 66,197 & Driver Distraction Detection \\ 
         \hline 
         \parbox[t]{4mm}{\multirow{2}{*}{\rotatebox[origin=c]{90}{\textbf{Temple}}}}
         &  TUAB~\cite{TUAB} & 256Hz & 16 & 10s & 409,455 & Abnormal EEG Classification\\
         &  TUEV~\cite{TUEV} & 256Hz & 16 & 5s & 112,464 & Event Detection  \\ \hline
    \end{tabular}
    \label{tbl:data}
    \vspace{-0.5cm}
\end{table}
\section{Experimental Results}

\begin{table*}[ht]
  \centering
  \setlength{\tabcolsep}{3pt}
  \setlength\extrarowheight{4pt}
  \small
  \caption{Performance of EEG2Rep in comparison to the competitors in a \textit{Linear Probing} setting.}
  \label{tab:performance_metrics}
  \begin{tabular}{lcccccccccccc}
    \toprule
    Models & \multicolumn{2}{c}{DREAMER} & \multicolumn{2}{c}{Crowdsourced} & \multicolumn{2}{c}{STEW} & \multicolumn{2}{c}{DriverDistraction} & \multicolumn{2}{c}{TUAB} & \multicolumn{2}{c}{TUEV} \\
    \cmidrule(lr){2-3} \cmidrule(lr){4-5} \cmidrule(lr){6-7} \cmidrule(lr){8-9} \cmidrule(lr){10-11} \cmidrule(lr){12-13}
    & Acc & AUROC & Acc & AUROC & Acc & AUROC & Acc & AUROC & Acc & AUROC & B-Acc & W-F1 \\
    \midrule
    BENDR~\cite{Bendr} & 51.48$\pm$\tiny{1.87} & 51.68$\pm$\tiny{1.98} & 70.46$\pm$\tiny{4.14} & 70.56$\pm$\tiny{4.32} & 63.03$\pm$\tiny{1.07} & 63.03$\pm$\tiny{1.07} & 68.40$\pm$\tiny{3.08} & 55.21$\pm$\tiny{3.17} & 72.78$\pm$\tiny{4.17} & 79.87$\pm$\tiny{4.11} & 37.38$\pm$\tiny{3.11} & 61.31$\pm$\tiny{3.18} \\
    MAEEG~\cite{maeeg} & \underline{54.24$\pm$\tiny{2.04}} & \underline{53.98$\pm$\tiny{2.68}} & 75.21$\pm$\tiny{2.11} & 75.01$\pm$\tiny{2.01} &  \underline{67.99$\pm$\tiny{1.86}} & \underline{68.58$\pm$\tiny{1.88}} & 68.37$\pm$\tiny{2.60} & 55.29$\pm$\tiny{2.20} & 72.62$\pm$\tiny{3.99} & 79.75$\pm$\tiny{4.07} & 37.23$\pm$\tiny{2.99} & 61.38$\pm$\tiny{3.08} \\
    TS-TCC~\cite{TS-TCC2021} & 53.60$\pm$\tiny{2.68} & 53.87$\pm$\tiny{2.28} & \underline{77.75$\pm$\tiny{2.94}} & \underline{77.83$\pm$\tiny{2.11}} & 64.54$\pm$\tiny{1.62} & 64.64$\pm$\tiny{1.73} & \underline{76.36$\pm$\tiny{3.48}} & 56.27$\pm$\tiny{2.88} & 74.39$\pm$\tiny{3.06} & 81.02$\pm$\tiny{2.97} & 35.98$\pm$\tiny{2.85} & 60.67$\pm$\tiny{2.98} \\
    TF-C~\cite{TF-C2022} & 52.52$\pm$\tiny{1.88} & 52.26$\pm$\tiny{2.08} & 64.82$\pm$\tiny{7.23} & 65.32$\pm$\tiny{6.12} & 58.84$\pm$\tiny{2.36} & 58.69$\pm$\tiny{2.43} & 64.85$\pm$\tiny{3.89} & 53.87$\pm$\tiny{4.02} & 69.33$\pm$\tiny{5.78} & 75.75$\pm$\tiny{3.75} & 30.12$\pm$\tiny{4.06} & 56.23$\pm$\tiny{4.05} \\
    BIOT~\cite{yang2023biot} & 53.35$\pm$\tiny{2.41} & 53.42$\pm$\tiny{1.91} & 76.23$\pm$\tiny{4.56} & 76.33$\pm$\tiny{4.12} &67.54$\pm$\tiny{2.08} & 67.69$\pm$\tiny{3.60} & 63.93$\pm$\tiny{1.28} & \underline{63.33$\pm$\tiny{1.25}} & \underline{75.11$\pm$\tiny{2.79}} & \underline{82.92$\pm$\tiny{2.01}} & \underline{40.02$\pm$\tiny{1.87}} & \underline{65.98$\pm$\tiny{2.01}} \\
    \rowcolor{blue!10} \textbf{EEG2Rep} & \textbf{58.45}$\pm$\tiny{1.82} & \textbf{55.19}$\pm$\tiny{1.95} & \textbf{81.66}$\pm$\tiny{2.93} & \textbf{81.67}$\pm$\tiny{2.65} & \textbf{69.04}$\pm$\tiny{1.04} & \textbf{69.10}$\pm$\tiny{1.23} & \textbf{76.88}$\pm$\tiny{2.55} & \textbf{65.59}$\pm$\tiny{2.43} & \textbf{76.55}$\pm$\tiny{3.33} & \textbf{83.24}$\pm$\tiny{3.25} & \textbf{43.25}$\pm$\tiny{3.12} & \textbf{69.95}$\pm$\tiny{3.21}
     \\
    \bottomrule
  \end{tabular}
  \label{tab:linear-prob}
\end{table*}

\subsection{Datasets}

We employed two distinct dataset types, totaling six datasets, to evaluate the effectiveness of our EEG2Rep model: i) Four datasets were obtained using different types of Emotiv headsets(14-channel wireless headsets capable of data collection in real-world scenarios). Three of these datasets are publicly available. The fourth dataset, named ``Driver Distraction'' is a private dataset provided by Emotiv, and collected using an older version of the headset~\footnote{https://www.emotiv.com/epoc/}. Reporting results on this private dataset along with other public ones allows us to assess our model's effectiveness across different Emotiv headset types, each producing specific types of noise. ii) We expanded our evaluation to another dataset type, the Temple University Hospital (TUH) Corpus~\cite{TUAB,TUEV}, collected in a different setting (in a laboratory setting using 24-36 channel braincaps). TUH is one of the largest open EEG data repositories, featuring diverse devices with varying channel numbers, all collected in clinical settings. An overview of the datasets is available in Table~\ref{tbl:data}, and additional descriptions, including details on data preprocessing, are provided in Appendix~\ref{app:data}.

To assess our model's performance, we partitioned the Emotiv datasets into subject-wise train/validation/test sets. This setup poses a challenge for models to learn generalized patterns, given the inter-subject variability. As for TUAB and TUEV, the training and test separation is inherent in the dataset. Additionally, we further split TUEV and TUAB training sets into 20\% validation and 80\% training subject-wise.
\vspace{-0.2cm}
\subsection{Competitors and Implementation Setup}
We conducted extensive comparisons against five state-of-the-art methods for EEG representation learning. These methods include invariance-based approaches TS-TCC~\cite{TS-TCC2021}, TF-C~\cite{TF-C2022}, and BIOT~\cite{yang2023biot} as well as self-prediction methods BENDR~\cite{Bendr} and MAEEG~\cite{maeeg}. To ensure a fair evaluation, we used publicly available code for the baseline methods. All experiments utilized the PyTorch framework on a system featuring a single Nvidia A5000 GPU (24GB). Model and hyperparameter selection relied on the validation set. Table~\ref{tab:linear-prob} and Table~\ref{tab:Fine-Tuning} present five sets of results with varied random seeds, reporting mean and standard deviation values. We adhere to the original VICReg~\cite{vicreg2022} for setting the hyperparameters $\lambda$, $\mu$, and $\gamma$ in the total loss. For further details on the evaluation metrics, refer to the Appendix~\ref{app:setting}.
\begin{table*}[htbp]
  \centering
  \setlength{\tabcolsep}{1.5pt}
  \setlength\extrarowheight{2pt}
  \small
  \caption{Performance of EEG2Rep in comparison to the competitors in a \textit{Fine Tuning} setting.}
  \label{tab:performance_metrics_2}
  \begin{tabular}{lcccccccccccc}
    \toprule

    \textbf{Models} & \multicolumn{2}{c}{DREAMER} & \multicolumn{2}{c}{Crowdsourced} & \multicolumn{2}{c}{STEW} & \multicolumn{2}{c}{DriverDistraction} & \multicolumn{2}{c}{TUAB} & \multicolumn{2}{c}{TUEV} \\
    \cmidrule(lr){2-3} \cmidrule(lr){4-5} \cmidrule(lr){6-7} \cmidrule(lr){8-9} \cmidrule(lr){10-11} \cmidrule(lr){12-13}
    & {Acc} & {AUROC} & {Acc} & {AUROC} & {Acc} & {AUROC} & {Acc} & {AUROC} & {Acc} & {AUROC} & {B-Acc} & {W-F1} \\
    \midrule
    BENDR~\cite{Bendr}   & 54.45$\pm$\tiny{2.11} & 53.02$\pm$\tiny{3.11} & 83.78$\pm$\tiny{2.35} & 83.8$\pm$\tiny{2.63} & 69.74$\pm$\tiny{2.11} & 69.77$\pm$\tiny{2.03} & 74.31$\pm$\tiny{2.38} & 59.86$\pm$\tiny{2.6} & 76.96$\pm$\tiny{3.98} & 83.97$\pm$\tiny{3.44} & 41.17$\pm$\tiny{2.89} & 67.31$\pm$\tiny{2.96} \\
    MAEEG~\cite{maeeg}   & 53.63$\pm$\tiny{2.61} & 52.08$\pm$\tiny{2.36} & 86.75$\pm$\tiny{3.50} & 86.21$\pm$\tiny{3.41} & \underline{72.46$\pm$\tiny{3.67}} & \underline{72.5$\pm$\tiny{3.22}} & \underline{74.58$\pm$\tiny{2.16}} & 60.79$\pm$\tiny{2.72} & 77.56$\pm$\tiny{3.56} & 86.56$\pm$\tiny{3.33} & 41.23$\pm$\tiny{3.65} & 67.38$\pm$\tiny{3.69} \\
    TS-TCC~\cite{TS-TCC2021}  & \underline{58.16$\pm$\tiny{2.11}} & \underline{55.05$\pm$\tiny{1.79}} & 89.22$\pm$\tiny{1.22} & 89.22$\pm$\tiny{1.22} & 71.00$\pm$\tiny{2.98} & 71.03$\pm$\tiny{3.02} & 74.21$\pm$\tiny{2.68} & 60.33$\pm$\tiny{2.66} & \underline{79.66$\pm$\tiny{2.99}} & 87.02$\pm$\tiny{2.68} & 40.98$\pm$\tiny{2.55} & 68.67$\pm$\tiny{2.89} \\
    TF-C~\cite{TF-C2022}    & 52.82$\pm$\tiny{1.66} & 52.86$\pm$\tiny{1.85} & 82.93$\pm$\tiny{4.02} & 82.90$\pm$\tiny{4.32} & 68.65$\pm$\tiny{1.12} & 68.70$\pm$\tiny{1.75} & 65.39$\pm$\tiny{4.12} & 58.75$\pm$\tiny{3.98} & 72.33$\pm$\tiny{5.64} & 78.48$\pm$\tiny{3.85} & 40.12$\pm$\tiny{3.66} & 66.23$\pm$\tiny{3.85} \\
    BIOT~\cite{yang2023biot}    & 53.45$\pm$\tiny{2.01} & 53.53$\pm$\tiny{2.13} & 87.95$\pm$\tiny{3.52} & 87.78$\pm$\tiny{3.09} & 69.88$\pm$\tiny{2.15} & 70.11$\pm$\tiny{2.57} & 74.34$\pm$\tiny{3.57} & \underline{61.21$\pm$\tiny{4.36}} & 79.21$\pm$\tiny{2.15} & \underline{87.42$\pm$\tiny{2.01}} & \underline{46.02$\pm$\tiny{1.68}} & \underline{69.98$\pm$\tiny{1.99}} \\
    EEG2Rep (Random) & 54.61$\pm$\tiny{2.22} & 53.61$\pm$\tiny{2.09} & \underline{91.19$\pm$\tiny{1.18}} & 91.22$\pm$\tiny{1.23} & 70.26$\pm$\tiny{1.59} & 70.49$\pm$\tiny{1.86} & 72.95$\pm$\tiny{2.95} & 59.5$\pm$\tiny{3.17} & 77.85$\pm$\tiny{3.14} & 84.91$\pm$\tiny{3.07} & 44.25$\pm$\tiny{3.01} & 68.95$\pm$\tiny{2.89} \\
    \rowcolor{blue!10} \textbf{EEG2Rep} (Pre-Trained) & \textbf{60.37}$\pm$\tiny{1.52} & \textbf{59.42}$\pm$\tiny{1.45} & \textbf{94.13}$\pm$\tiny{2.11} & \textbf{94.13}$\pm$\tiny{2.17} & \textbf{73.60}$\pm$\tiny{1.47} & \textbf{74.40}$\pm$\tiny{1.50} & \textbf{80.07}$\pm$\tiny{2.63} & \textbf{66.14}$\pm$\tiny{2.44} & \textbf{80.52}$\pm$\tiny{2.22} & \textbf{88.43}$\pm$\tiny{3.09} & \textbf{52.95}$\pm$\tiny{1.58} & \textbf{75.08}$\pm$\tiny{1.21}

 \\
    \bottomrule
  \end{tabular}
  \label{tab:Fine-Tuning}
  \vspace{-0.2cm}
\end{table*}
\subsection{Linear Probing}
Table \ref{tab:linear-prob} presents the average performance of EEG2Rep along with other state-of-the-art methods over five runs. For each dataset, the number in \textbf{bold} indicates the highest accuracy achieved, while the number \underline{underlined} represents the second best (This formatting is consistent across all tables presented in this paper). The results presented in Table \ref{tab:linear-prob} indicate that our model, EEG2Rep, achieves the highest average performance on all EEG tasks. 

The DREAMER and STEW datasets exhibit high inter-subject variance, stemming partly from the limited number of patients in the recordings and partly due to the complexity of the EEG tasks. As observed in the results, self-prediction-based methods like MAEEG and EEG2Rep tend to capture more general patterns that can be applied across different subjects.  
In contrast, tasks such as eyes open/eyes close in Crowdsourced dataset are easier to generalize among subjects due to the nature of the EEG task. TUAB also experiences low inter-subject variance as it encompasses a substantial number of subjects (more than 1000 patients). The results highlight that invariance-based methods like TS-TCC and BIOT outperform BENDER and MAEEG in Crowdsourced and TUAB. However, EEG2Rep while being a self-prediction technique, manages to improve the semantic level of representations resulting in the best average accuracy among all competitors. 

\begin{table}[ht]
\centering
\setlength{\tabcolsep}{0.5pt}
\setlength\extrarowheight{2pt}
\small
\caption{EEG2Rep model performance in cross-domain settings on STEW and Crowdsourced datasets.}
  \vspace{-0.3cm}
\begin{tabular}{lcccc}
\hline
\multirow{2}{*}{\textbf{Models}}
 & \multicolumn{2}{c}{STEW} & \multicolumn{2}{c}{Crowdsourced} \\ \cline{2-5} 
                & ACC & AUROC & ACC & AUROC \\ \hline
Random initialization & 70.26$\pm$\tiny{1.59} & 70.49$\pm$\tiny{1.86} & 91.19$\pm$\tiny{1.18} & 91.22$\pm$\tiny{1.23} \\ 
In-domain pre-trained & 73.60$\pm$\tiny{1.47} & 74.40$\pm$\tiny{1.50} & \textbf{94.13$\pm$\tiny{2.11}} & \textbf{94.13$\pm$\tiny{2.17}} \\ 
Pre-trained on DREAMER & \underline{73.75$\pm$\tiny{1.95}} & \underline{74.95$\pm$\tiny{2.77}} & 93.91$\pm$\tiny{1.78} & 93.86$\pm$\tiny{1.80} \\
Pre-trained on DriverDistraction & 73.68$\pm$\tiny{2.17} & 74.67$\pm$\tiny{3.06} & \underline{94.09$\pm$\tiny{1.95}} & \underline{94.11$\pm$\tiny{2.02}} \\ 
Pre-trained on All Emotiv & \textbf{74.11$\pm$\tiny{2.34}} & \textbf{77.38$\pm$\tiny{2.77}} & 94.05$\pm$\tiny{1.68} & 94.07$\pm$\tiny{1.75} \\ \hline
\end{tabular}
\label{tab:cross-domain}
  \vspace{-0.3cm}
\end{table}

\vspace{-0.2cm}
\subsection{Fine Tuning}
Table \ref{tab:Fine-Tuning} presents the average classification accuracy results across different datasets, comparing the performance of EEG2Rep with other pre-trained models that were initialized using their respective pretext tasks. These results are consistent with the ones presented in Table~\ref{tab:linear-prob}, and our EEG2Rep achieves the highest average performance on all EEG tasks. 

The comparison also includes another version of the EEG2Rep model, initialized randomly and denoted as EEG2Rep (Random). This is equivalent to supervised training. The table shows that using pre-trained EEG2Rep leads to an average accuracy improvement of 6\% on average compared to EEG2Rep (Random). 
Significant enhancements are evident in specific datasets, particularly DREAMER, DriverDistraction, and TUEV. In DREAMER and DriverDistraction, EEG2Rep demonstrates AUROC improvement of 5.81\% and 2.26\%, respectively, when compared to random initialization. Similarly, for TUEV, there is a 6.85\% improvement in weighted F1, validating the effectiveness of incorporating informative context input in self-supervised methods for enhanced learning and improved EEG classification. It's worth noting that in datasets with pronounced subject invariance issues, such as DREAMER, STEW, Driverdistraction, and TUAB, the performance of the supervised models is notably subpar. In some cases, these models even exhibit lower performance compared to models pre-trained self-supervised and utilized for classification through a linear layer. For instance, in DREAMER, linear probed EEG2Rep achieves a 3.84\% higher accuracy than its supervised counterpart.

\vspace{-0.3cm}
\subsection{Cross-Domain}
We evaluated the performance of our EEG2Rep model in a cross-domain setting, where the model is trained on one dataset and tested on another. This approach assesses the model's ability to generalize across different types of EEG domains and tasks. We utilized two Emotiv datasets for the target tasks: STEW and Crowdsourced, both characterized by limited training samples, making cross-domain evaluation particularly relevant. 

As depicted in Table~\ref{tab:cross-domain}, pre-training the model with datasets like DREAMER led to performance improvements for the STEW dataset, where the task is mental workload classification. This improvement can be attributed to the contextual alignment between activities recorded in the DREAMER and STEW datasets. Interestingly, even though the DriverDistraction dataset is not directly related to mental workload classification, the experimental environment and certain distraction classes are similar to those in the STEW dataset. This similarity allowed the EEG2Rep model to benefit from pre-training on the DriverDistraction dataset, as reflected in the performance.

For the Crowdsourced dataset, where the task is eye open/closed classification, the model achieved robust performance when pre-trained on all available Emotiv datasets. However, this performance did not surpass that achieved through in-domain pre-training. This outcome suggests that the alignment of brain activities across datasets is critical for optimal performance and indicates the potential need to utilize a larger number of pre-training datasets to cover various subjects and tasks comprehensively.

\begin{table}[ht]
  \centering
  \setlength{\tabcolsep}{10pt}
  \setlength\extrarowheight{2pt}
  \caption{Ablation Study of EEG2Rep Components}
    \vspace{-0.3cm}
\begin{tabular}{lc} \hline
 \rowcolor{blue!10} \textbf{EEG2Rep} & \textbf{Average Accuracy: 67.64} \\ \hline \hline
\textbf{Masking Strategy} & \\ \hline
Block Masking & 66.45 \textcolor{red}{($\downarrow$ 1.19)}\\
Random Masking & 60.19 \textcolor{red}{($\downarrow$ 7.45)}\\ \hline \hline

\textbf{Targets} & \\ \hline
Input-Space Prediction & 54.52 \textcolor{red}{($\downarrow$ 13.12)} \\ \hline
\textbf{Loss Component} & \\ \hline \hline
W/O Variance-Covariance & 65.63 \textcolor{red}{($\downarrow$ 2)}\\ \hline
\end{tabular}
\label{tab:ablation}
\vspace{-0.3cm}
\end{table}

\vspace{-0.3cm}
\subsection{Ablation Study}
\subsubsection*{\textbf{Masking Strategy}}
We compare our semantic subsequence preserving (SSP) strategy with other random and block masking strategies. For random masking, we adopt an approach similar to masking autoencoders~\cite{mae2022}, where patches are shuffled randomly, and the initial $50\%$ of these patches are selected. In the case of block masking, we follow the MAEEG~\cite{maeeg} approach, emphasizing the masking of a continuous chunk.
The results in Table~\ref{tab:ablation} highlight the effectiveness of semantic subsequence preservation in guiding our model towards learning meaningful representations. The subpar performance in the random masking strategy could be attributed to the model primarily attempting to interpolate the masked values rather than focusing on learning a semantic representation. With block masking, the visible subsequence may have a shorter length than the natural time scale of the brain, posing a significant challenge to the reconstruction process.

\subsubsection*{\textbf{Masking Ratio}}
\begin{table*}[ht]
\centering
\setlength{\tabcolsep}{5pt}
\setlength\extrarowheight{2pt}
\caption{Performance comparison between random masking and SSP across different datasets and masking ratios.}
\begin{tabular}{ccccccccccccc}
\hline
\multirow{3}{*}{\textbf{Mask Ratio}} & \multicolumn{2}{c}{DREAMER} & \multicolumn{2}{c}{Crowdsourced} & \multicolumn{2}{c}{STEW} & \multicolumn{2}{c}{DriverDistraction} & \multicolumn{2}{c}{TUAB} & \multicolumn{2}{c}{TUEV} \\ \cline{2-13} 
                             & Random & SSP & Random & SSP & Random & SSP & Random & SSP & Random & SSP & Random & SSP \\ \hline
 \cellcolor{blue!10} \textbf{10\%}& 52.11  & 53.02 & 72.13  & 76.69 & 63.13  & 68.14 & 64.02  & 70.41 & 64.50  & 72.36 & 33.01  & 40.37 \\ 
 \cellcolor{blue!10}\textbf{25\%} & \textbf{53.23}  & 54.73 & 75.05  & 77.28 & \textbf{65.12}  & \textbf{70.47} & \textbf{65.40}  & 72.94 & 64.75  & 73.73 & 33.14  & 42.53 \\ 
 \cellcolor{blue!10} \textbf{50\%}& 52.63  & \textbf{59.89} & \textbf{76.69}  & \textbf{81.24} & 64.97  & 69.52 & 64.12  & \textbf{76.12} & \textbf{68.12}  & \textbf{76.55} & \textbf{34.59}  & \textbf{44.23} \\
 \cellcolor{blue!10} \textbf{75\%} & 51.77  & 52.03 & 72.10  & 76.24 & 63.27  & 64.69 & 63.12  & 65.03 & 66.34  & 70.81 & 35.28  & 42.19 \\ 
 \cellcolor{blue!10}\textbf{90\%} & 51.51  & 51.95 & 70.77  & 73.44 & 62.05  & 63.70 & 62.06  & 64.91 & 66.34  & 70.12 & 34.14  & 38.72 \\ \hline
 \cellcolor{blue!10} \textbf{Average}& 52.25  & \textbf{54.32} & 73.35  & \textbf{76.98} & 63.71  & \textbf{67.30} & 63.74  & \textbf{69.88} & 66.01  & \textbf{72.71} & 34.03  & \textbf{41.61} \\ \hline
\end{tabular}
\label{tab:masking}
  \vspace{-0.25cm}
\end{table*}
The masking ratio during self-supervised pretraining is crucial as it determines both the difficulty of the self-prediction task and the quality of the learned representations. Therefore, identifying an optimal masking ratio is essential for effective representation learning from EEG data. Following the recent study on the effects of masking on representation learning~\cite{kong2023understanding}, we pre-train our model using five different mask rates [10\%, 25\%, 50\%, 75\%, 90\%] with our SSP method and random masking to evaluate the quality of representation learning. As shown in Table~\ref{tab:masking}, several interesting findings emerged from the results. We observed that conservative masking (10\%) leads to low performance, regardless of the masking strategy. However, as the masking ratio increases, the performance of the model improves, indicating that higher masking can result in more high-level representation learning. Nevertheless, performance degradation occurs with overly aggressive masking (90\%), suggesting that representation does not become monotonically more high-level with increasing masking aggressiveness. In other words, overly aggressive masking also leads to low-level representations, similar to conservative masking. Moderate and high masking ratios yield the best results for both random and SSP masking strategies.

As shown in Table~\ref{tab:masking}, our SSP masking strategy consistently achieves higher performance than random masking across various masking ratios. A key distinction between our proposed masking strategy in EEG2Rep and random masking lies in the arrangement of visible patches. In random masking, there is a possibility that visible patches are not consecutive, whereas in EEG2Rep, visible patches (referred to as preserved patches) are ensured to remain contiguous. Additionally, during the reconstruction of masked patches near the boundary between masked and unmasked regions, the model uses nearby visible patches to interpolate, thereby capturing low-level information, which resonates with the empirical observations in~\cite{mae2022,kong2023understanding}. Our proposed masking strategy is less susceptible to such phenomena compared to random masking, as it ensures that masked tokens remain contiguous. The high number of boundaries in random masking may result in less contextualized learning. Our empirical results further confirm that our proposed method yields higher performance compared to random masking, even with aggressive and conservative masking.

\subsubsection*{\textbf{Masking Blocks}}
\begin{figure}
    \centering
    \includegraphics[trim=0cm 0cm 0cm 0cm, width=0.48\textwidth]{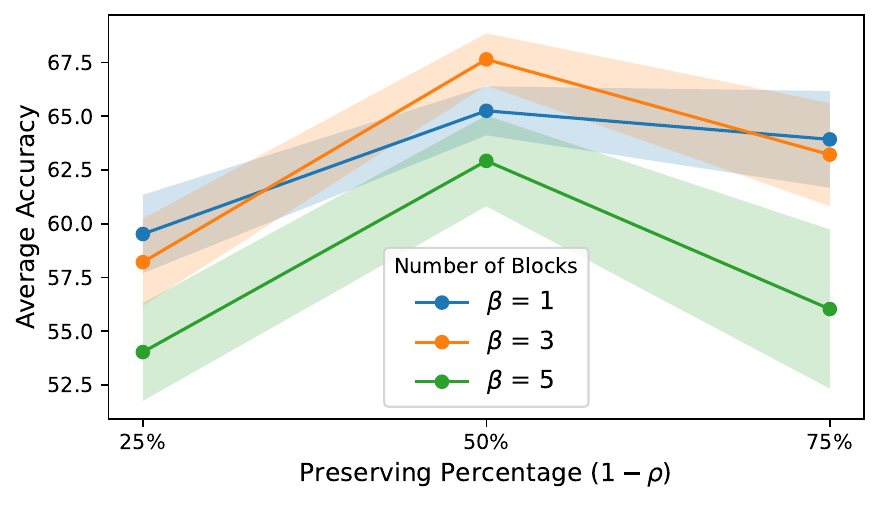}
    \caption{Effect of preserving percentage ($1-\rho$) on average accuracy across all EEG datasets: A comparison of accuracy variation across different numbers of blocks ($\beta$), with error bars indicating standard deviation.}
    \label{fig:Masking-ratio}
      \vspace{-0.3cm}
\end{figure} 
In EEG2Rep, our focus is on preserving subsequences to ensure information is retained in the masked input. Figure~\ref{fig:Masking-ratio} depicts the average accuracy of EEG2Rep relative to the number of masked blocks and the ratio of preserved subsequences (inverse of masking). Optimal results are observed when preserving 50\% of the input in 3 blocks and masking the other half. When preserving only 75\% of the input, it is more effective to mask a single chunk rather than multiple chunks, as the model tends to interpolate instead of learning semantic patterns in EEG data. 

\subsubsection*{\textbf{Input-Space Prediction}}
Table \ref{tab:ablation} shows a comparison of linear probing performance when the loss is computed in input-space versus representation space. We hypothesize that a key component of EEG2Rep lies in computing the loss entirely in the representation space, enabling the target encoder to generate abstract prediction targets that eliminate irrelevant raw details. As evident from Table \ref{tab:ablation}, predicting in input-space leads to a $13.12\%$ degradation in linear probing performance, highlighting the impact of noise and wide range of amplitudes inherent to EEG data.

\subsubsection*{\textbf{Loss Regularization}}

As discussed in Section~\ref{sec:LossReg}, the self-supervised loss function comprises two components: reconstruction and regularization. The main role of the reconstruction term is to ensure similarity between the representations of masked and unmasked versions of the same sample. In contrast, the regularization term not only aids in preventing representation collapse but also enhances the representation learning task. As shown in Table~\ref{tab:ablation}, the regularization term contributes to a 2\% accuracy improvement on average across all EEG tasks.

\subsubsection*{\textbf{Robustness to Noise}}
\begin{figure}
    \centering
    \includegraphics[trim=0cm 0cm 0cm 0cm, width=0.46\textwidth]{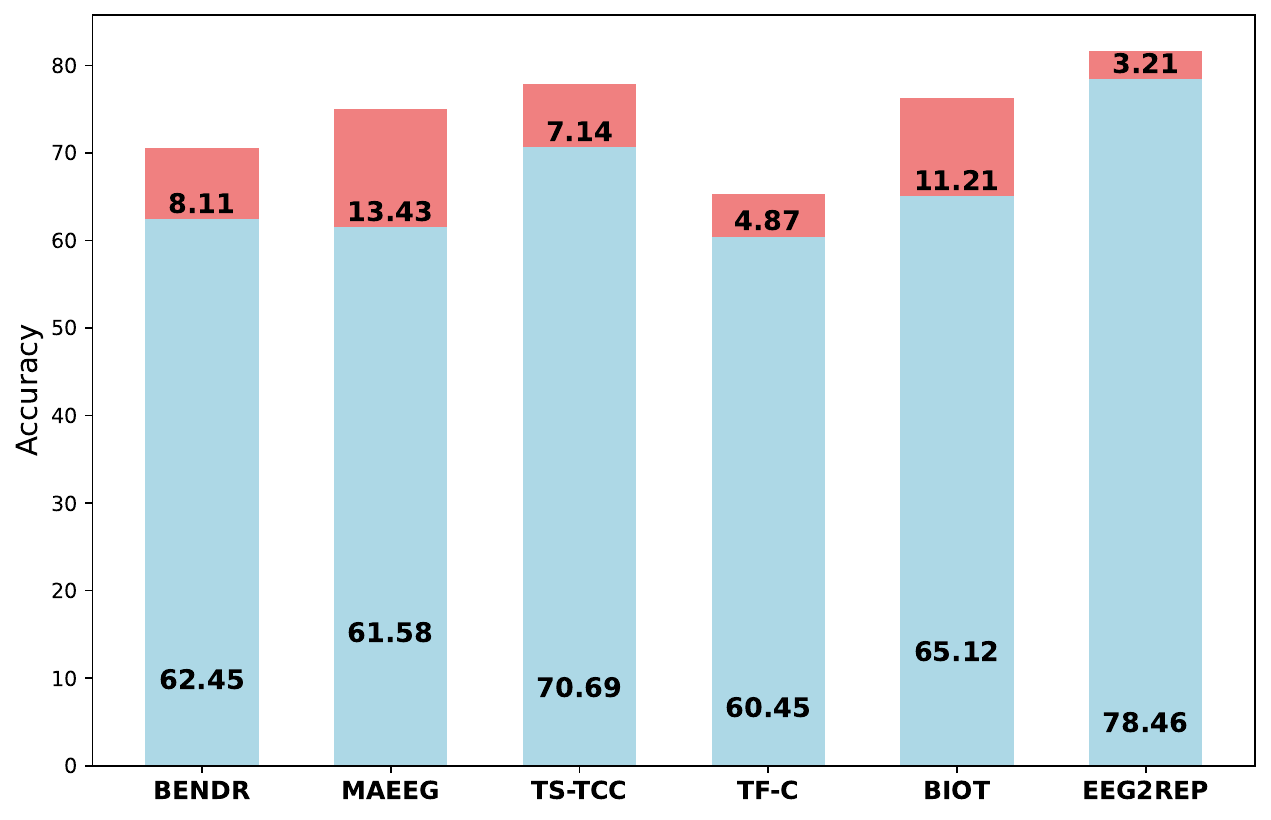}
      \vspace{-0.2cm}
    \caption{Model Robustness Comparison: Assessing model performance by introducing Gaussian noise, DC-shift, and amplitude changes to Crowdsourced data.}
    \label{fig:Robustness}
      \vspace{-0.5cm}
\end{figure} 

In this experiment, our goal is to assess the robustness of our model to noise, recognizing the high signal-to-noise ratio as a significant challenge in learning representations from EEG data. To do so we introduce Gaussian noise, time shift, DC-shift, and amplitude scaling, as recommended by neurologists~\cite{MohsenvandIM20} to the Crowdsourced dataset. Such transformation do not change the interpretation of the EEG according to~\cite{MohsenvandIM20}. The details of noise types are described in appendix~\ref{app:noise}. As depicted in Figure~\ref{fig:Robustness}, our model demonstrates superior robustness to these noises in the data, while raw input-based models like BENDER and MAEEG exhibit the most degradation in accuracy. TS-TCC and TF-C show commendable robustness to these noises, possibly due to their utilization of similar augmentations for their invariance losses.

\section{Conclusion}

EEG2Rep emerges as a pioneering self-supervised approach tailored to address the inherent challenges of EEG data representation learning, including low signal-to-noise ratio. By innovatively predicting masked inputs in the latent representation space and employing a novel semantic subsequence preserving method, EEG2Rep facilitates the generation of rich semantic representations. Our extensive experiments across six diverse EEG tasks demonstrate that EEG2Rep not only significantly surpasses state-of-the-art methods but also highlights remarkable robustness to noise. We found that preserving 50\% of EEG recordings optimizes accuracy across all tasks. In the future, we will explore the alignment of this finding with the natural time scale of the brain on different tasks.
\bibliographystyle{IEEEtran}
\bibliography{Reference}

\appendix
\section{Dataset Overview and Processing} \label{app:data}
\subsection{\textbf{Emotiv}}
We applied a bandpass filter to all Emotiv datasets and windowed them into segments consisting of 256 time steps, each equivalent to 2 seconds of recording.

\vspace{0.2cm}
\noindent\textbf{DREAMER} \\

DREAMER is a multimodal database featuring electroencephalogram (EEG) and electrocardiogram (ECG) signals recorded during affect elicitation using audio-visual stimuli~\cite{dreamer} using a 14-channel Emotiv EPOC headset. The dataset includes signals from 23 participants, accompanied by their self-assessments of affective states after each stimulus in terms of valence, arousal, and dominance. For our classification task, we specifically utilize the arousal labels. The DREAMER dataset can be obtained from here\footnote{\url{https://zenodo.org/records/546113}}, and we employ the Torcheeg toolkit for preprocessing, which involves cropping and applying low-pass and high-pass filters\footnote{\url{https://torcheeg.readthedocs.io/en/v1.1.0/torcheeg.datasets.html}}. It is important to note that, for our analysis, we solely focus on EEG data, and the ECG signals are excluded from consideration.

\vspace{0.2cm}
\noindent\textbf{Crowdsourced} \\
Crowdsourced EEG data was collected while participants were engaged in a resting state task, involving periods with eyes open and eyes closed, each lasting 2 minutes. Out of 60 participants, only 13 successfully completed both conditions using 14-channel EPOC+, EPOC X, and EPOC devices. The data was initially recorded at 2048 Hz and later downsampled to 128 Hz. Raw EEG data for these 13 participants, along with preprocessing, analysis, and visualization scripts, are openly available on the Open Science Framework (OSF)\footnote{\href{https://osf.io/9bvgh/?view_only=70744f62157c46d5bd731480db1873df}{https://osf.io/9bvgh}}.

\vspace{0.2cm}
\noindent\textbf{Simultaneous Task EEG Workload (STEW)} \\ 
STEW dataset comprises raw EEG recordings from 48 participants using a 14-channel Emotiv EPOC headset involved in a multitasking workload experiment utilizing the SIMKAP multitasking test~\cite{stew}. Additionally, the subjects' baseline brain activity at rest was recorded before the test. The data was captured using the Emotiv Epoc device with a sampling frequency of 128Hz and 14 channels, resulting in 2.5 minutes of EEG recording for each case. Participants were instructed to assess their perceived mental workload after each stage using a rating scale ranging from 1 to 9, and these ratings are available in a separate file. Moreover, this dataset includes binary class labels, considering a workload rating of more than 4 as high and otherwise as low. We utilize these labels for our specific problem. STEW can be accessed upon request through the IEEE DataPort\footnote{\url{https://ieee-dataport.org/open-access/stew-simultaneous-task-eeg-workload-dataset}}.

\vspace{0.2cm}
\noindent\textbf{DriverDistraction} \\ 

The data were gathered by recording the EEG brain signals of 17 participants, each using a driving simulator for approximately 40 minutes.
The participants performed various distraction activities while they were driving. These can be grouped into the following high-level activities: 1. Talking to a passenger 2. Using a phone (texting and calling) 3. Problem solving. The EEG data were sampled at a frequency of 128Hz, through 14 channels on the Emotiv Epoc EEG headset. The sampling result is a multivariate (14 input variables) time-series dataset containing approximately 5.5 million records. The data were then manually annotated with the activity being performed at each time point.

\subsection{Temple University Hospital (TUH) EEG Corpus}

\vspace{0.2cm}
\noindent\textbf{TUH Abnormal EEG Corpus (TUAB)} \\
The TUH Abnormal EEG Corpus (TUAB) is a subset of the Temple University Hospital (TUH) EEG Corpus, which is one of the largest publicly available collections of clinical EEG data. The TUAB specifically focuses on EEG recordings labeled as abnormal, making it a valuable resource for studies on neurological disorders, brain function anomalies, and the development of diagnostic tools~\cite{TUAB}.

\vspace{0.2cm}
\noindent\textbf{TUH EEG Events (TUEV)} \\
The TUH EEG Events Corpus (TUEV) contains annotations of EEG segments classified into six different categories: spike and sharp wave, generalized periodic epileptiform discharges, periodic lateralized epileptiform discharges, eye movement, artifact, and background~\cite{TUEV}.

\vspace{0.2cm}
\noindent\textbf{Acquisition and Preprocessing}\\
The TUH Abnormal EEG Corpus (TUAB)~\cite{TUAB} and TUH EEG Events (TUEV)~\cite{TUEV} can be accessed upon request through the Temple University Electroencephalography (EEG) Resources\footnote{\url{https://isip.piconepress.com/projects/tuh_eeg/html/downloads.shtml}}. We processed both datasets to adhere to the 16 EEG montages~\cite{yang2023biot}, following the 10-20 international system, are as follows: "FP1-F7", "F7-T7", "T7-P7", "P7-O1", "FP2-F8", "F8-T8", "T8-P8", "P8-O2", "FP1-F3", "F3-C3", "C3-P3", "P3-O1", "FP2-F4", "F4-C4", "C4-P4", and "P4-O2".

\section{Details of Experimental Settings} \label{app:setting}
\subsection{Parameter Setting}
In our experiment, the EEG2Rep model employed one depth-wise and two spatial-wise convolution layers for input embedding, each with 16 filters. During training, a batch size of 256 was used, and we utilized the Adam optimization algorithm \cite{kingma2014adam}. To prevent overfitting, we implemented an early stopping method based on the validation loss. The model was pre-trained for 500 epochs, after which logistic regression was applied to the representations for linear probing. Similar to the transformer-based model for multivariate time series classification (TST) \cite{TST2021} and the default transformer block \cite{attention2017}, in our experiments, we employed eight attention heads in both the context and target encoder to capture diverse features from the EEG. The transformer encoding dimension was set to $d_e = 16$, and the feed-forward network (FFN) in the transformer block expanded the input size by a factor of 4 before projecting it back to its original size. For the learning rate, we start with $1 \times 10^{-3}$ and use a cosine learning rate scheduler to adjust it over time \cite{data2vec2022}.

\subsection{Noise Types for Model Robustness}\label{app:noise}
Table~\ref{tab:noise} provides the details of the noise types we added to the Crowdsourced dataset to test the robustness of EEG2Rep and benchmark models to noise. 

\begin{table}[h]
\caption{Noise Types Details}
\centering
\begin{tabular}{lc|c}
\hline
\textbf{Transformation} & \textbf{Min} & \textbf{Max} \\
\hline
Amplitude Scale & 0.5 & 2 \\
Time Shift & -50 & 50 \\
DC Shift & -10 & 10 \\
Additive Gaussian Noise & 0 & 0.2 \\
\hline
\end{tabular}
\label{tab:noise}
\end{table}

\subsection{\textbf{Evaluation Metrics}}
\label{app:metrics}
\textbf{(Balanced) Accuracy} is defined as the average recall obtained for each class. We use the term \textit{ACC} for binary classification and \textit{B-ACC} for multi-class classification. \textbf{AUROC} represents the area under the ROC curve, condensing the ROC curve into a single number that measures the model's performance across multiple thresholds. It is employed for binary classification. \textbf{Weighted F1} is utilized for multi-class classification in this paper. It is a weighted average of individual F1 scores for each class, with each score weighted by the number of samples in the corresponding class.

\end{document}